\documentclass[aip,reprint,jcp]{revtex4-1}

\usepackage{graphicx}
\usepackage{dcolumn}
\usepackage{epstopdf}
\usepackage{hyperref}

\begin{document}

\title{Predicting polymorphism in molecular crystals using orientational entropy}

\author{Pablo M. Piaggi}
\affiliation{Theory and Simulation of Materials (THEOS), {\'E}cole Polytechnique F{\'e}d{\'e}rale de Lausanne (EPFL),  CH-1015  Lausanne,  Switzerland}
\affiliation{Facolt{\`a} di Informatica, Instituto di Scienze Computazionali, and National Center for Computational Design and Discovery of Novel Materials (MARVEL), Universit{\`a} della Svizzera italiana (USI), Via Giuseppe Buffi 13, CH-6900, Lugano, Switzerland}
\author{Michele Parrinello}%
\email{parrinello@phys.chem.ethz.ch}
\affiliation{Department of Chemistry and Applied Biosciences, ETH Zurich, c/o USI Campus, Via Giuseppe Buffi 13, CH-6900, Lugano, Switzerland}
\affiliation{Facolt{\`a} di Informatica, Instituto di Scienze Computazionali, and National Center for Computational Design and Discovery of Novel Materials (MARVEL), Universit{\`a} della Svizzera italiana (USI), Via Giuseppe Buffi 13, CH-6900, Lugano, Switzerland}

\date{\today}

\begin{abstract}
We introduce a computational method to discover polymorphs in molecular crystals at finite temperature.
The method is based on reproducing the crystallization process starting from the liquid and letting the system discover the relevant polymorphs. 
This idea, however, conflicts with the fact that crystallization has a time scale much longer than that of molecular simulations. 
In order to bring the process within affordable simulation time, we enhance the fluctuations of a collective variable by constructing a bias potential with well tempered metadynamics.
We use as collective variable an entropy surrogate based on an extended pair correlation function that includes the correlation between the orientation of pairs of molecules. 
We also propose a similarity metric between configurations based on the extended pair correlation function and a generalized Kullback-Leibler divergence.
In this way, we automatically classify the configurations as belonging to a given polymorph using our metric and a hierarchical clustering algorithm.
We find all relevant polymorphs for both substances and we predict new polymorphs.
One of them is stabilized at finite temperature by entropic effects.
\end{abstract}

\maketitle

Polymorphism is the ability that substances have to crystallize into different structures.
A paradigmatic example is carbon that in its two main polymorphs, graphite and diamond, exhibits amazingly different properties.
Polymorphism is also important from a practical point of view since controlling which crystal structure forms is of the utmost importance in many manufacturing processes. 
The pharmaceutical industry suffers in particular the consequences of polymorphism\cite{Bernstein02,Hilfiker06}.
Active pharmaceutical ingredients are usually small, organic molecules that frequently exist in a plethora of crystalline forms.
Different polymorphs can be patented separately and usually lead to different drug performances.
Therefore a comprehensive screening of polymorphs is crucial to avoid a rival company from releasing to the market the same molecule in a different polymorph\cite{Cabri07}, and to anticipate the transformation of one polymorph into another during the manufacturing process or the shelf life\cite{Bauer01}.  

The screening of polymorphs was traditionally performed experimentally in spite of the large costs involved\cite{Hilfiker06}. 
In the last 15 years the increase in computer power and the development of algorithms able to screen a large number of polymorphs has lead to a very significant successes in polymorph prediction\cite{Bazterra02,Oganov06,Price08,Pickard11,Yu11}. 
Such methods are based on the search of local minima on the potential energy surface. 
The minima are ordered by energy and typically corrected for thermal effects using the harmonic approximation.
However no method can claim to be able to scan exhaustively all the relevant low-lying minima. 
In addition, entropic effects beyond the harmonic approximation can be significant. 
Not only they can alter the delicate energetic balance between the different polymorphs but even stabilize structures that are not local minima of the potential energy surface. 
Another issue that is often overlooked is the kinetic side of crystallization, for instance a given polymorph can be favored relative to energetically lower ones by the fact that is kinetically more accessible.
For all these reasons we take here a different approach and we try to reproduce on the computer the crystallization process starting from the liquid state and letting the system discover all the relevant polymorphs. 

This ambition conflicts with the fact that crystallization is a process that occurs on a time scale that is much longer than that of computer simulations.  
This requires the use of enhanced sampling methods that bring the time scale of crystallization within affordable simulation time\cite{Valsson16review}. 
Some enhanced sampling methods require the definition of order parameters or collective variables.
This methods channel and enhance the fluctuations so as to favor the reversible observation of multiple freezing and melting events.  
Thus far such order parameters have been based on some structural geometrical information on the phase the system is going to crystallize into. 
If one is interested in discovering new polymorphs this approach defeats the purpose.  
Recently, however, we have shown that in simple systems this can be circumvented by using as collective variable surrogates of enthalpy and entropy\cite{Piaggi17}.
The idea is to mimic what happens in a real system in which there is a trade off between entropy and enthalpy.  
We dealt with simple one component\cite{Piaggi17} or two component\cite{Mendels18} atomic systems. 
In this case the following expression as a surrogate for entropy:
\begin{equation}
        S_2=-2\pi\rho k_B \int\limits_0^{\infty} \left [ g(r) \ln g(r) - g(r) + 1 \right ] r^2 dr ,
        \label{eq:pair_entropy}
\end{equation}
where $r$ is a distance, $g(r)$ is the radial distribution function and $\rho$ is the density of the system was used.
Together with enthalpy, $S_2$ proved successful in predicting the lattice into which the system was going to crystallize.  
For a discussion of $S_2$ we refer the reader to Ref. \citenum{Piaggi17}. 
This has been a simple proof of principle to show that crystal structures, even the ones that are stabilized by entropy\cite{Mendels18}, can be predicted.  

Molecular systems that are of interest to pharmaceutical industry present a complexity much larger than the relatively simple systems so far handled in which most of the times only one polymorph was stable. 
Here we enlarge considerably the scope of these calculations and move to study molecular systems that, as we shall see, present a large number of polymorphs.

We shall consider a system of molecules and, for the purpose of developing a collective variable, we shall represent each molecule by the position of its center of mass and a vector that characterizes its orientation in space.
We can define a correlation function $g(r,\theta)$ akin to $g(r)$ but including the relative orientation between two molecules.
$\theta$ is defined as $\theta=\arccos \left ( \frac{\mathbf{v}_i \cdot \mathbf{v}_j}{ |\mathbf{v}_i||\mathbf{v}_j|} \right )$ where $\mathbf{v}_i$ and $\mathbf{v}_j$ are the vectors characterizing the orientation of molecules $i$ and $j$.
Statistical mechanics provides us with an expression for the entropy of such a system equivalent to the one ini Eq.~\ref{eq:pair_entropy}, this is\cite{Prestipino04},
\begin{eqnarray}
        S_{\theta}=-\pi\rho k_B \int\limits_0^{\infty} \int\limits_0^{\pi} [ & g(r,\theta) & \ln g(r,\theta) \nonumber \\
                                                                            -& g(r,\theta) & + 1  ] r^2 \sin \theta  \: dr \: d\theta .
        \label{eq:pair_orientational_entropy}
\end{eqnarray}
We shall use $S_{\theta}$ as collective variable to drive simulations.
A similar collective variable was introduced in ref. \cite{Gobbo18} although in that case the probability as a function of the angle of the molecules with respect to a fix reference frame was used to define the entropy.

An important part in the definition of our CV is the choice of angles to characterize the relative orientation between neighboring molecules.
In principle, three angles are needed to specify completely the relative orientation between two rigid molecules, for instance the three Euler angles $\phi,\theta,\psi$.
This would imply the construction of a function $g(r,\phi,\theta,\psi)$ whose calculation would be cumbersome.
Here we take a different approach and we use several CVs each involving one angle.

We shall choose two systems to test the ability of $S_{\theta}$ to explore polymorphism, namely urea and naphthalene.
We have chosen two CVs and therefore two angles for each system.
In the case of urea we use the angles $\theta_1$ and $\theta_2$ to define the CVs $s_{\theta_1}$ and $s_{\theta_2}$.
The first one is defined using the direction of the dipole moment and the second with the direction of the vector joining the two nitrogens.
In the case of naphthalene we use the direction of the longest axis of the molecule and the direction perpendicular to the aromatic rings to define the CVs $s_{\theta_2}$ and $s_{\theta_1}$.

\begin{figure}
        \begin{center}
                \includegraphics[width=0.99\columnwidth]{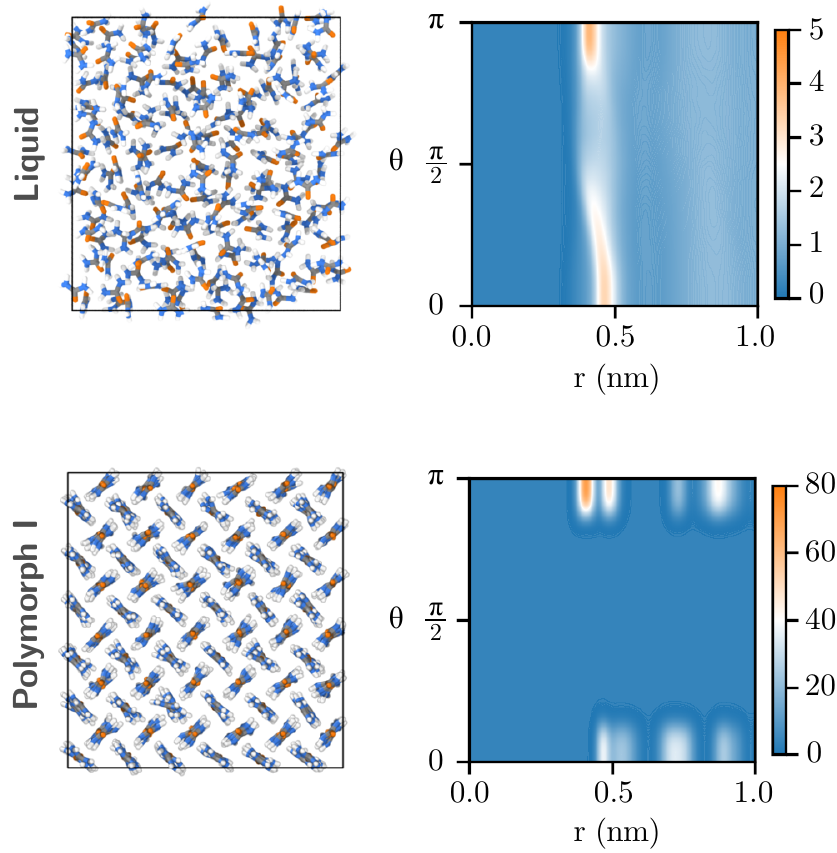}
                \caption{\label{fig:Figure1}
		$g(r,\theta)$ for the liquid and polymorph I of urea at 450 K.
		Snapshots of the system in each of the phases are shown.
    }
        \end{center}
\end{figure}

Since the use of $g(r,\theta)$ is not so widespread, we thought useful to help the reader get a feeling of its behavior by plotting $g(r, \theta_1)$ for the liquid and polymorph I of urea at 450 K (see Fig.\ \ref{fig:Figure1}).
The liquid $g(r,\theta_1)$ exhibits some structure at very short distances and almost no correlations at distances larger than 0.8 nm. 
On the other hand the $g(r,\theta_1)$ of polymorph I shows a well defined structure that persists at long distances as expected from a solid phase.
As can be observed in the figure, one of the main characteristic of polymorph I is that molecules have parallel or antiparallel dipole moments.
Thus, $g(r,\theta)$ contains important orientational information that can help to distinguish between phases.

We briefly describe the polymorphs found experimentally so far for each system.
Urea shows a rich polymorphism and up to five polymorphs have been reported\cite{Lamelas05,Olejniczak09,Dziubek17}.
The most stable form at ambient conditions is form I and it has been extensively studied.
Other two forms exist at higher pressures, namely forms III and IV. 
Another high pressure polymorph, form V, has been found although to our knowledge the details of the structure have not been reported.
There has also been theoretical work that found other polymorphs\cite{Giberti15,Shang17}.
In particular, for urea as described by the Amber force field, the so called form A \cite{Giberti15,Shang17} is highly relevant having an energy very close to that of the ground state.
At variance with urea, naphtalene has only one solid form and in spite of several investigations at high pressure\cite{Fabbiani06,Likhacheva14} no new forms have yet been found.

We have used well-tempered metadynamics (WTMetaD) \cite{Barducci08} to enhance the fluctuations of $s_{\theta_1}$ and $s_{\theta_2}$.
In WTMetaD a time-dependent potential is constructed as a sum of kernels, typically chosen to be Gaussians.
The potential discourages frequently visited configurations thus boosting the exploration of configuration space.
Further details can be found in the Materials and Methods section.
In the 200 ns biased simulations both urea and naphthalene explore thoroughly the space spanned by the CVs, although understanding the nature of the configurations explored requires further analysis. 
A visual inspection of the trajectories shows many transitions to different crystal forms.
The crystalline configurations have different orientations in space and some of them contain small crystalline defects.
The wealth of information that these simulations contain, however, cannot be analyzed with the naked eye.
It would therefore be useful to have an automatic method to identify and classify the polymorphs that crystallize in the course of the simulation.
In the following paragraphs we propose one such automatic method.

A key ingredient for an automatic method to identify and classify polymorphs is a metric for the similarity between two given configurations.
Several structural similarity metrics exist in the literature\cite{De16} but in this work we shall propose a new one.
In the present context, it is natural to use for this purpose the very function $g(r,\theta)$ that defines the CVs to characterize the configuration of the system.
However, we still need a measure of distance between two $g(r,\theta)$.
We can define a distance taking inspiration in the pair entropy expression.
We first note that Eq.~\ref{eq:pair_orientational_entropy} is a measure of the distance between the $g(r,\theta)$ of the present configuration and the $g(r,\theta)$ of the ideal gas, i.e. $g(r,\theta) = 1 \: \forall \: r,\theta $.
Inspired by this observation we introduce a divergence of $g_1(r,\theta)$ with respect to $g_2(r,\theta)$,

\begin{widetext}
\begin{equation}
        D(g_1 || g_2) =\int\limits_0^{\infty} \int\limits_0^{\pi} \left [ g_1(r,\theta) \ln \frac{g_1(r,\theta)}{g_2(r,\theta)} - g_1(r,\theta) + g_2(r,\theta) \right ] r^2 \sin \theta  \: dr \: d\theta.
        \label{eq:divergence}
\end{equation}
\end{widetext}
This is a generalization of the Kullbak-Leibler divergence for non-normalized functions. 
This divergence is a special case of Bregman divergence and has some interesting properties such as that of being convex and having a minimum at $g_1=g_2$ \cite{Cesa06}. 
Strictly speaking $D(g_1 || g_2)$ is not a distance since it is not symmetric.
For applications in which a well defined distance is needed we shall use a symmetrized version of Eq.~\ref{eq:divergence}, namely,
\begin{equation}
        d(g_1,g_2) = \frac{D(g_1 || g_2)+D(g_2 || g_1)  }{2}.
        \label{eq:distance}
\end{equation}

\begin{figure*}[ht!]
        \begin{center}
                \includegraphics[width=0.99\textwidth]{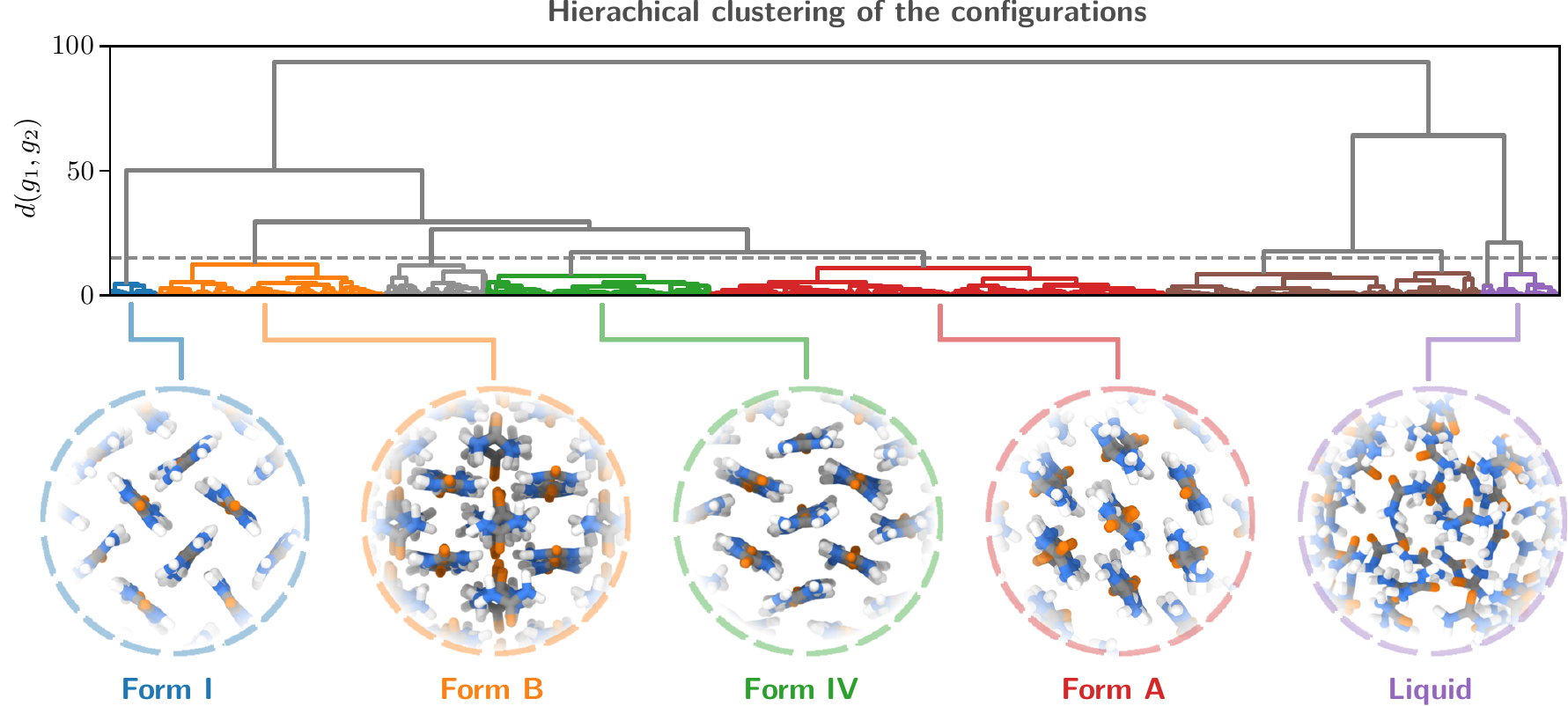}
                \caption{\label{fig:Figure2} Tree diagram resulting from the clustering according to the distance in Eq.~\ref{eq:distance} of the trajectory of urea at 450 K. The threshold distance used to join clusters is shown with a grey dashed line. Configurations at 450 K for selected clusters are shown.
    }
        \end{center}
\end{figure*}

Equipped with this metric, we can compare configurations and analyze the rich and complex trajectories resulting from the biased simulations.
We will exemplify our approach by analyzing the trajectory of urea.
The configurations in the trajectory were clustered using a hierarchical clustering approach\cite{Mullner13,Jones14} based on the distance defined in Eq.~\ref{eq:distance}.
We used the average distance between points in two clusters as linkage criterion.
As a result of the clustering, we obtain a tree diagram (see Fig.\  \ref{fig:Figure2}) that shows the similarity between different configurations in the trajectory.
We can now choose a threshold distance $d_c$ and join together all configurations that belong to a branch with maximum distance $d_c$ between configurations.
The choice of $d_c$ allows us to focus on the dominant structures that appear in the simulation.
In Fig.\ \ref{fig:Figure2} $d_c$ is shown with a dashed line and the resulting clusters are shown with different colors.

We still have to determine the structures that each cluster represents.
A possible way to do so is by choosing the minimum energy configuration within each cluster.
This configuration will be the one with the least number of defects and less affected by the thermal motion of molecules.
In some cases this approach is not appropriate, for instance when structures are stabilized by large entropic effects.
In these cases one can choose the configuration that has an energy close to the average energy of the cluster.
We have chosen with this criteria the configurations that are used to determine the nature of each cluster.
Some of these configurations are shown in Fig.\ \ref{fig:Figure2}.

We now describe the phases that were found.
The tree diagram has two main branches.
The right branch contains liquid-like configurations (violet cluster in Fig.\ \ref{fig:Figure2}) and interesting partially ordered configurations (brown cluster in Fig.\ \ref{fig:Figure2}) in which the dipole moments are oriented in the same direction but do not exhibit long range translational order.
The left branch contains solid-like configurations and it can be further subdivided into five relevant clusters.
One of these clusters contains an unstable structure and we shall disregard it (grey cluster in Fig.\ \ref{fig:Figure2}).
The other four clusters correspond to form I, to a new polymorph that we shall name form B, to form IV and form A.
To the best of our knowledge it is the first time that form B has been reported.
The other structures were expected based on previous studies \cite{Olejniczak09,Giberti15,Shang17}.
All polymorphs are metastable at 450 K and they do not transform during a 1 ns unbiased simulation.
We include the configurations of all relevant structures in the SI.
We have also performed a similar analysis for naphthalene.
The clustering identifies the experimentally known form I, the liquid, and a new structure that we shall name form A.
The results can be found in the SI.

\begin{figure}[b]
        \begin{center}
                \includegraphics[width=0.99\columnwidth]{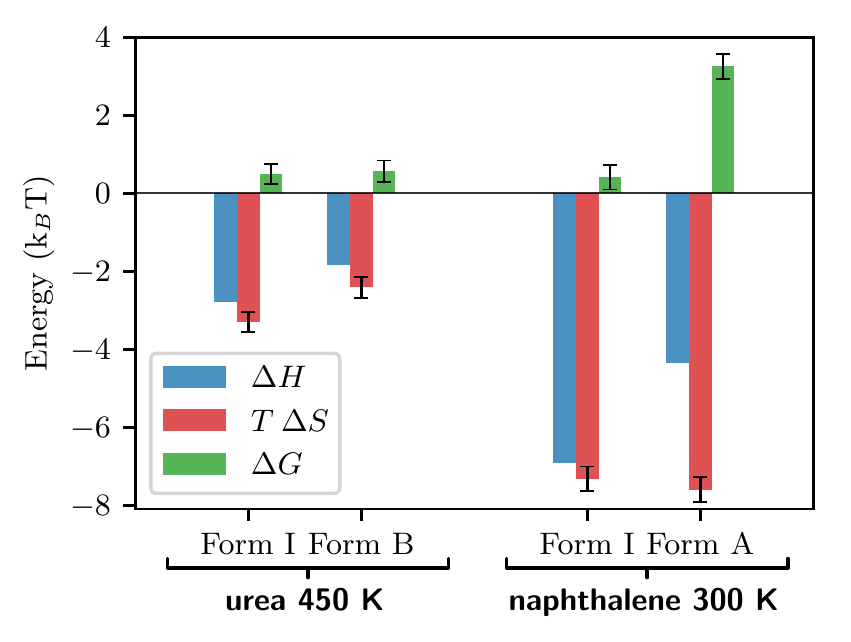}
                \caption{\label{fig:Figure3} Enthalpy, entropy and free energy for selected polymorphs of urea at 450 K and naphthalene at 300 K. All quantities have the liquid as reference state.
    }
        \end{center}
\end{figure}

We have estimated the free energy difference between the polymorphs and the liquid using:
\begin{equation}
  \Delta G = -\frac{1}{\beta} \log \left ( \frac{p_i}{p_l} \right )
\end{equation}
where $p_i$ and $p_l$ the probabilities to observe polymorph $i$ and the liquid, respectively.
In an unbiased MD simulation one could calculate the probabilties $p_i$ and $p_l$ directly from the simulation.
However, since we have introduced the WTMetaD potential that alters the probability of observing a given configuration, the $p_i$'s must be calculated with the reweighting procedure described in ref. \citenum{Tiwary14}.
We have employed the clustering described above to identify the phase of each configuration.
The resulting free energy differences are shown in Fig.\ \ref{fig:Figure3}.
The error bars in this figure are relatively large since free energy differences calculated in this way are not easy to converge and the simulation contains transitions between many different structures.
Fig.\ \ref{fig:Figure3} shows also the enthalpy $\Delta H$ of the polymorphs with respect to the liquid phase.
Using $\Delta G$ and $\Delta H$ the entropy $\Delta S$ can be calculated from the definition of free energy $\Delta G=\Delta H-T\: \Delta S$.
The results show that form I of urea is close to equilibrium with the liquid at 450 K, in line with ref. \citenum{Salvalaglio12} and \citenum{Giberti15} where the melting temperature was found to be around 420 K.
Similarly, form I of naphthalene is close to equilibrium with the liquid at 300 K, as expected from the estimated melting temperature (330 K).

\begin{figure}
        \begin{center}
                \includegraphics[width=0.99\columnwidth]{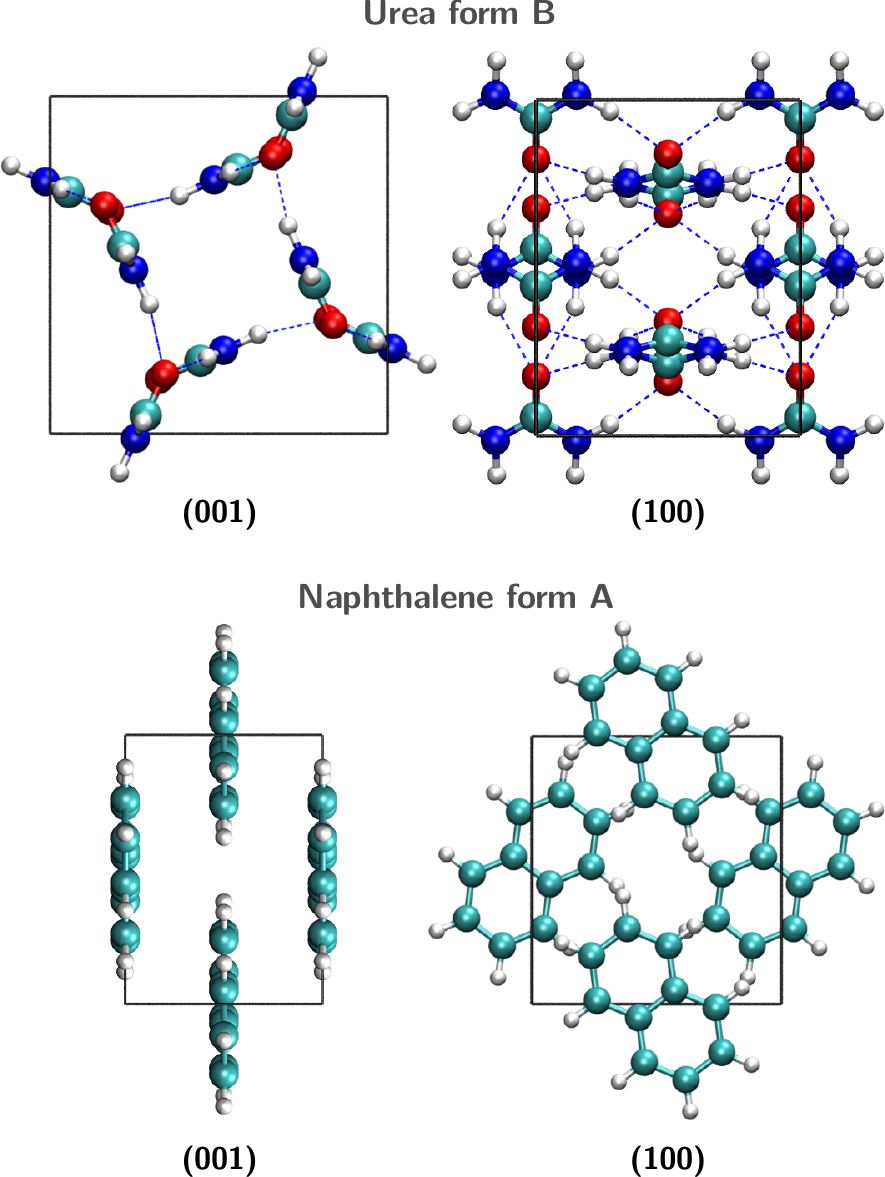}
                \caption{\label{fig:Figure4} Crystal structures of the new forms of urea and naphthalene. Images obtained with VMD\cite{Humphrey96}.
    }
        \end{center}
\end{figure}

We shall now consider in detail the newly discovered polymorphs.
We first discuss form B of urea that has a P4$_2$/mbc space group and is shown in Fig.\ \ref{fig:Figure4}.
This polymorph is particularly interesting because it has a relatively high enthalpy, roughly k$_B$T above form I (see Fig.\ \ref{fig:Figure3}).
Based only on energy arguments one would conclude that this structure cannot compete with form I.
However, strong entropic effects stabilize it.
The entropies shown in Fig.\ \ref{fig:Figure3} indeed show a greater contribution to the stability in form B than in form I.
We suggest that an important factor that contributes to the entropy is the fast rotation about the C-O axis.
We have calculated the characteristic rotation time using the time autocorrelation function of the N-N unit vector and fitting an exponential function to it.
We show the results in Fig.\ 2 of the SI and we compare them with those of form I.
The characteristic time of rotation in form I is $\sim 800$ ps while in form B it is $\sim 7$ ps.
We have also computed the probability $p(\theta)$ as a function of the rotation angle $\theta$ about the C-O axis.
From $p(\theta)$ the free energy can be calculated as $G(\theta)=-k_B T \log p(\theta) \sin \theta$.
We show the results in Fig.\ 2 of the SI.
Both in form I and B $G(\theta)$ exhibits a barrier separating two molecular configurations in which the N are exchanged.
The barrier height is $\sim 18$ kJ/mol in form B while it is $\sim 34$ kJ/mol in form I.
The entropy contribution from this rotation can be calculated from $k_B T \int p(\theta)\:\log p(\theta)  \sin \theta d\theta$.
The difference in entropy between form I and B accounts for about 1.5 kJ/mol (0.4 k$_B$T).
As the temperature is lowered, the structure undergoes a phase transition at around 200 K.
Therefore, methods that search structures at zero temperature would only find the low temperature form instead of the high temperature one.
The change in structure cannot be accounted for using harmonic corrections.

We now turn to discuss polymorph A of naphthalene.
Form A has a layered structure and its space group is Pnnm\cite{Isotropy}.
The structure is shown in Fig.\ \ref{fig:Figure4}.
During an unbiased simulation at 300 K, form A decays to the liquid.
This is consistent with the calculated free energy (see Fig.\ \ref{fig:Figure3}) that shows that form A has a free energy around 3 $k_BT$ higher than form I and the liquid. 
In spite of the relatively high free energy, it is possible that this polymorph could be kinetically trapped.

We have presented a method to explore polymorphism in molecular crystals in finite temperature molecular dynamics simulations.
An important feature of our method is that not only does it discover polymorphs but also pinpoints which are the relevant ones at a given thermodynamical condition.
In fact, the new polymorph of urea, form B, could have not been predicted from a zero temperature search with harmonic corrections.
A key ingredient of our approach is the structure similarity metric defined using $g(r,\theta)$ and the new distance in Eq.~\ref{eq:divergence}.
This metric allows us to automatically assign configurations to a given polymorph, thus reducing the burden of the analysis of the simulations.
We are also able to calculate free energies and entropies from the simulation using a reweighting procedure\cite{Tiwary14}.
In the future, we plan to generalize our approach to crystals with more complex hydrogen bond networks and to molecules with internal degrees of freedom.


\section*{Materials and Methods}

Urea and naphthalene were described using the generalized amber force field (GAFF) \cite{Wang04}.
For naphthalene, the electrostatic potential was calculated at the B3LYP/6-31+G(d,p) level using Gaussian 09\cite{Frisch09} and the partial charges of the atoms were fitted using the restrained electrostatic potential (RESP) method\cite{Bayly93}.
The partial charges of urea were those provided with the Amber 03 database\cite{Case05}.
Biased MD simulations were performed using Gromacs 5.1.4 \cite{Abraham15} patched with a development version of PLUMED 2 \cite{Tribello14}.
Van der Waals interactions and the electrostatic interaction in real space were calculated with cutoffs 0.9 nm and 0.75 nm for urea and naphthalene, respectively.
The electrostatic interaction in reciprocal space was calculated using the particle mesh ewald (PME) method \cite{Essmann95}.
The atomic bonds involving hydrogen were constrained using the LINCS algorithm\cite{Hess08} and the equations of motion were integrated with a 2 fs timestep.
The temperature was controlled using the stochastic velocity rescaling thermostat \cite{Bussi07} with a relaxation time of 0.1 ps.
We mantained the pressure at its atmoshperic value employing the isotropic version of the Parrinello-Rahman \cite{Parrinello81} barostat with a 10 ps relaxation time.
We employed systems of 108 and 36 molecules for urea and naphtalene, respectively. 

We now provide the parameters used for the WTMetaD simulations\cite{Barducci08}.
The Gaussians had a width of 0.1 k$_B$ and 0.2 k$_B$ for urea and naphthalene, respectively.
In all cases the Gaussians had a height of 5 k$_B$T and were deposited every 1 ps.
The bias factor was 200 for all simulations.

We now discuss some practical aspects of the use of $S_{\theta}$ as a CV.
In order to calculate the forces arising from the WTMetaD bias, $S_{\theta}$ should be continuous and differentiable.
This can be achieved by constructing the function $g(r,\theta)$ using Gaussian kernels of width $\sigma_r$ and $\sigma_{\theta}$, as done in previous work.
Furthermore, the integration in Eq.~\ref{eq:pair_orientational_entropy} cannot have an infinite upper limit, and in practice a finite cutoff $r_m$ is taken.
The integration is performed numerically using the trapezoid rule with steps of size $\sigma_r$ and $\sigma_{\theta}$ in the $r$ and $\theta$ dimension, respectively.
We report in Table \ref{tab:table1} the chosen parameters.
$\sigma_{\theta}$ is reported in units of $\cos{\theta}$.
A subtlety in the calculation of $S_{\theta}$ is the periodicity of $g(r,\theta)$ in its $\theta$ argument.
For a general molecule $g(r,\theta)$ is periodic in $\theta$ with period $\pi$.
However, for a molecule that has a mirror symmetry with respect to the plane perpendicular to the vector $v$ defining the orientation of the molecule,  $g(r,\theta)$ has a period $\pi/2$.
We report in Table  \ref{tab:table1} whether a given CV is defined based on a direction of the molecule with mirror symmetry.

\begin{table}[t]
\centering
\caption{\label{tab:table1}%
 Parameters used in the definition of CVs $S_{\theta_1}$ and $S_{\theta_2}$ for urea and naphthalene. See text for details.}
\begin{ruledtabular}
\begin{tabular}{ccccc}
    & $r_m$ (nm) & $\sigma_r$ (nm) & $\sigma_{\theta}$ & mirror symmetry \\ 
\hline
 \multicolumn{5}{c}{urea} \\ 
 $S_{\theta_1}$ & 0.6 & 0.05 & 0.25 & no \\ 
 $S_{\theta_2}$ & 0.6 & 0.05 & 0.125 & yes \\ 
\hline
 \multicolumn{5}{c}{naphtalene} \\ 
 $S_{\theta_1}$ & 0.7 & 0.05 & 0.125 & yes \\ 
 $S_{\theta_2}$ & 0.7 & 0.05 & 0.125 & yes \\ 
\end{tabular}
\end{ruledtabular}
\end{table}

\begin{acknowledgments}
We are grateful to Zoran Bjelobrk for providing the force field for naphthalene.
We would also like to thank Haiyang Niu for his valuable assistance in the analysis using the structure factor.
This research was supported by the NCCR MARVEL, funded by the Swiss National Science Foundation.
The authors also acknowledge funding from European Union Grant No. ERC-2014-AdG-670227/VARMET. 
The computational time for this work was provided by the Swiss National Supercomputing Center (CSCS) under Project ID mr3. 
Calculations were performed in CSCS cluster Piz Daint.
\end{acknowledgments}

\section*{References}

\end{document}